# A bivariate marginal likelihood specification of spatial econometric modeling of very large datasets

Giuseppe Arbia[*]

*Department of Statistics, Università Cattolica del Sacro Cuore, Rome, 00198 Italy.*

______________________________________________________________________

**A B S T R A C T**

______________________________________________________________________

This paper proposes a bivariate marginal likelihood specification of spatial econometrics models that simplifies the derivation of the log-likelihood and leads to a closed form expression for the estimation of the parameters. With respect to the more traditional specifications of spatial autoregressive models, our method avoids the arbitrariness of the specification of a weight matrix, presents analytical and computational advantages and provides interesting interpretative insights. We establish small sample and asymptotic properties of the estimators and we derive the associated Fisher information matrix needed in confidence interval estimation and hypothesis testing.

.______________________________________________________________________

## 1. Introduction

Cross sectional linear spatial regressions have been based for almost four decades on the Cliff and Ord (1973) paradigm of autoregressive models which (founding on the seminal contribution of Whittle, 1954) represent the most popular way of taking into account the violation of the ideal conditions of independence among regression disturbances (Anselin, 1988, Arbia, 2006; LeSage and Pace, 2010). A popular estimation strategy for these models is based on (quasi) ML estimators assuming normality of the innovations, a procedure that ensures the asymptotic optimality properties (Lee, 2004). However, in practical instances, the likelihood function associated to the Cliff-Ord type models cannot be maximized analytically due to high degree of non-linearity in the parameters. It can indeed, be maximized numerically, but the computational procedures can be highly demanding both in terms of computing time and of the required computer storage (even with the current computational power!) when n is very large, as it more and more happens in many empirical applications. A very popular alternative to ML is the GMM approach suggested by Kelejian and Prucha (1998), a procedure that is computationally simple even in large samples. Within the context of a likelihood approach, LeSage and Pace (2007) suggested a matrix exponential spatial specification (MESS) of spatial models, a method based on the procedure introduced by Chiu et al. (1996) for covariance matrix modeling. The MESS approach is theoretically very simple, suggests a closed form solution for the estimators and improves dramatically the numerical performances with respect to the conventional spatial autoregressive models. Our contribution proceeds along the same direction indicated by LeSage and Pace (2007). Following Arbia (2012) we suggest an alternative specification of the spatial dependence based on the joint bivariate modeling of the disturbance terms and we propose the use of partial maximum likelihood techniques in the estimation phase. A similar approach was suggested by Wang *et al.* (2013) in the context of spatial discrete choices modeling. In the present paper we show that this approach produces significant simplifications dramatically reducing the computational burden and allows insights in the interpretation of the parameters and of the associated spatial spillovers. The approach does not require a full specification of a weights matrix and the estimation phase is computationally very simple in that a closed-form solution of the ML estimators can be obtained. We also establish the statistical properties of the suggested estimators proving unbiasedness and normality, deriving their asymptotic properties and the associated Fisher information matrix needed in confidence interval estimation and hypothesis testing. The core of the contribution is contained in Section 2 while Section 3 contains some comments on the interpretation of the parameters showing the advantages of the proposed approach and Section 4 concludes with some directions of further developments in this field.

---

[*] Tel: +39 06 3015 6109. *E-mail address*: giuseppe.arbia@rm.unicatt.it

**2. The model**

Given n observations related to a scalar random variable $y_i$ ($i = 1,...,n$ being spatial units) and of a non-stochastic random variable $x_i$, let us consider the following linear regression model:

$$y_i = \beta x_i + \varepsilon_i \qquad (1)$$

where $\varepsilon_i$ is the unobservable disturbance in location *i*. Both x and y are assumed to be centered on their respective expected values. We introduce the following assumption.

***Assumption 1.*** The regression disturbances follow a joint bivariate Gaussian distribution:

$$\begin{pmatrix} \varepsilon_i \\ \varepsilon_l \end{pmatrix} \approx MVN\left(0\mathbf{1}, \sigma^2 \Omega\right) \quad \forall i, l \quad l \in N(i) \qquad (2)$$

where $N(i)$ represents the set of neighbors of location *i*, however defined, $\Omega = \begin{pmatrix} 1 & \psi \\ \psi & 1 \end{pmatrix}$ is the correlation matrix and $\psi$ is the parameter controlling for the error spatial correlation such that $\psi \in [-1; +1]$. The assumption of normality is not essential to the method and can be easily relaxed by considering other bivariate density functions. The hypothesis that the correlation between units in a neighborhood is the same may seem restrictive, but in fact it derives from the implicit assumption of isotropy that is usually made in the literature (Arbia, 2006). If no directional bias occurs (anisotropy) then rather than considering a lagged value as the average over all neighbors, we can restrict, without loss of generality, to only one single neighbor randomly chosen. Specified in this way, Equations (1) and (2) represent the counterpart of the popular Spatial Error Model (Arbia, 2006).

Before introducing a partial *ML* procedure for the estimation of the unknown parameters of model (1) – (2), let us first introduce the definition of a *bivariate coding* for the available observations extending the work of Besag (1972, 1974).

**Definition 1:** *Just for the purpose of illustration, let us assume that the n observations are available on an regular square lattice grid and let us label the interior cells of such a grid with a cross "x" as indicated in Figure 1.*

*Figure 1: Bivariate coding pattern in a regular square lattice grid*

In particular let us code with a cross $q < n$ ($i=1,..., q$) of the *n* available spatial observations, ($q \subset Q$), let us also define $N(i)$ as the set of neighbors of location i, and let us further code with a cross *q* further locations ($l=1,..., q$) chosen randomly in the neighborhood of location *i*. The disturbance $\varepsilon_i$ and $\varepsilon_l$, $l \in N(i)$, are assumed to be spatially dependent due to their proximity while the pairs $\{\varepsilon_i, \varepsilon_l\}$ and $\{\varepsilon_j, \varepsilon_k\}$ (with $k \in N(j)$) are assumed to be stochastically independent provided $j, k \notin N(i, l)$, with $N(i, l) = \{N(i) \cup N(l)\}$ the joint neighborhood of *i* and *l*. Similar coding schemes can be easily introduced in irregular spatial schemes with an appropriate definition of neighborhood. By coding the units in this way, we are able to retain the spatial information contained in the sample by selecting pairs of neighboring spatial units without incurring in the problems of simultaneity typical of spatial econometrics models by selecting pairs of spatial units that are, by definition, independent on one another. The hypothesis of independent pairs, that may seem too restrictive at a first sight, has to be considered with reference to the definition of neighborhood. In fact we can always define a neighborhood which is large enough to create a buffer zone around each of the selected pair so as to produced independence among them. Finally considering pairs of random disturbances is not essential to the method and one could equally consider triplets and higher order groups, the only theoretical justification for this restriction being the Hammersey-Clifford theorem (Hammersey and Clifford, 1971) and the consequent restriction to pairwise interaction assumed by the auto-models definition of random fields (Besag, 1974). The above definition will be referred to as a *bivariate coding pattern*.

Under *Assumption 1* and *Definition 1*, we have that:

$$f_{\varepsilon_i \varepsilon_l}(\varepsilon_i, \varepsilon_l) = \frac{1}{2\pi\sigma^2\sqrt{1-\psi^2}} \exp\left\{-\frac{1}{2\sigma^2(1-\psi^2)}\left[\varepsilon_i^2 - 2\psi\varepsilon_i\varepsilon_l + \varepsilon_l^2\right]\right\} \text{ if } l \in N(i), \qquad (3)$$

The above bivariate density is characterized by the vector of parameters $\theta \equiv (\beta, \sigma^2, \psi)$. As a consequence of Equation (3) the (pseudo) likelihood associated to the *2q* selected units can be expressed as follows:

$$L(\theta) = L(\beta, \sigma^2, \psi) = \prod_{i=1}^{q} f_{\varepsilon_i \varepsilon_l}(\varepsilon_i, \varepsilon_l) = \prod_{i=1}^{q} \frac{1}{2\pi\sigma^2\sqrt{1-\psi^2}} \exp\left\{-\frac{1}{2\sigma^2(1-\psi^2)}\left[\varepsilon_i^2 - 2\psi\varepsilon_i\varepsilon_l + \varepsilon_l^2\right]\right\} =$$

$$= (2\pi)^{-q}(\sigma^2)^{-q}(1-\psi^2)^{-\frac{q}{2}} \exp\left\{-\frac{1}{2\sigma^2(1-\psi^2)}\sum_{i=1}^{q}\left[\varepsilon_i^2 - 2\psi\varepsilon_i\varepsilon_l + \varepsilon_l^2\right]\right\} = \qquad (4)$$

and correspondently the log-likelihood as:

$$l(\theta) = l(\beta, \sigma^2, \psi) = -q\ln(2\pi) - q\ln(\sigma^2) - \frac{q}{2}\ln(1-\psi^2) - \frac{1}{2\sigma^2(1-\psi^2)}\sum_{i=1}^{q}\left[\varepsilon_i^2 - 2\psi\varepsilon_i\varepsilon_l + \varepsilon_l^2\right] \qquad (5)$$

Such a likelihood falls within the general class of *pseudo-likelihood* (Pace and Salvan, 1997) and in the literature is sometimes referred to as *partial* (Cox, 1975) or *composite* likelihood (Lindsay, 1988; Varin *et al.*, 2011). A similar idea is contained in the contributions of Cox and Reid (2004) who defined a *pairwise likelihood* based on marginal bivariate events related to pairs of observations and of Nott and Ryden (1999) in the specific context of image analysis. Let us now introduce the following definitions.

**Definition 2:** Let $\alpha_1, \alpha_2, \alpha_3, \alpha_4, \alpha_5$ and $\alpha_6$, being a set of sufficient statistics defined as follows:

$$\alpha_1 = \sum_{i=1}^{q} x_i^2 + \sum_{l=1}^{q} x_l^2 = \sum_{j=1}^{2q} x_j^2 \qquad \alpha_2 = \sum_{i=1}^{q} y_i^2 + \sum_{l=1}^{q} y_l^2 = \sum_{j=1}^{2q} y_j^2$$

$$\alpha_3 = \sum_{i=1}^{q} x_i y_i + \sum_{l=1}^{q} x_l y_l = \sum_{j=1}^{2q} x_j y_j \qquad \alpha_4 = \sum_{i=1}^{q} x_i y_l + \sum_{l=1}^{q} x_l y_i \qquad (6)$$

$$\alpha_5 = \sum_{i=1}^{q} x_i x_l \qquad \alpha_6 = \sum_{i=1}^{q} y_i y_l$$

**Theorem 1:** Under Assumption 1, Definition 1 and Definition 2, the bivariate marginal maximum likelihood estimators (BML) of the vector of parameters $\theta = (\beta, \psi, \sigma^2)$ are given by the solution of the system of equations:

$$\begin{cases} \hat{\beta}_{BML} = \dfrac{\hat{\psi}_{BML}\alpha_4 - \alpha_3}{2\hat{\psi}_{BML}\alpha_5 - \alpha_1} \\[2ex] \hat{\sigma}^2_{BML} = \dfrac{\alpha_2 + \hat{\beta}^2_{BML}\alpha_1 - 2\hat{\beta}_{BML}\alpha_3 - 2\hat{\psi}_{BML}\alpha_6 - 2\hat{\psi}_{BML}\hat{\beta}^2_{BML}\alpha_5 + 2\hat{\psi}_{BML}\hat{\beta}_{BML}\alpha_4}{2q(1-\hat{\psi}^2)} \\[2ex] \hat{\psi}_{BML} = \dfrac{\alpha_6 - \hat{\beta}_{BML}\alpha_4 + \hat{\beta}^2_{BML}\alpha_5}{q\hat{\sigma}^2_{BML}} \end{cases} \qquad (7)$$

*Proof:* Let us consider the first-order conditions for the derivation of the bivariate marginal maximum likelihood estimators. First of all from Equation (5) we have:

$$\frac{\partial}{\partial \beta}l(\beta, \sigma^2, \psi) = \frac{\partial}{\partial \beta}\left\{const - q\ln(\sigma^2) - \frac{q}{2}\ln(1-\psi^2) - \frac{\sum_{i=1}^{q}\varepsilon_i^2}{2\sigma^2(1-\psi^2)} + \frac{2\psi\sum_{i=1}^{q}\varepsilon_i\varepsilon_l}{2\sigma^2(1-\psi^2)} - \frac{\sum_{l=1}^{q}\varepsilon_l^2}{2\sigma^2(1-\psi^2)}\right\} =$$

or, using Equation (1)

$$= \frac{1}{2\sigma^2(1-\psi^2)} \frac{\partial}{\partial \beta} \left\{ -\sum_{i=1}^{q}(y_i - \beta x_i)^2 + 2\psi \sum_{i=1}^{q}(y_i - \beta x_i)(y_l - \beta x_l) - \sum_{l=1}^{q}(y_l - \beta x_l)^2 \right\}$$

$$= \frac{1}{2\sigma^2(1-\psi^2)} \left[ 2\sum_{i=1}^{q}(y_i - \beta x_i)x_i + \left\{ 2\psi \frac{\partial}{\partial \beta}\left[ \sum_{i=1}^{q} y_i y_l - \beta \sum_{i=1}^{q} x_i y_l - \beta \sum_{i=1}^{q} x_l y_i + \beta^2 \sum_{i=1}^{q} x_i x_l \right] \right\} + 2\sum_{l=1}^{q}(y_l - \beta x_l)x_l \right]$$

$$= \frac{1}{\sigma^2(1-\psi^2)} \left[ \sum_{i=1}^{q}(y_i - \beta x_i)x_i + \left\{ \psi \frac{\partial}{\partial \beta}\left[ \sum_{i=1}^{q} y_i y_l - \beta \sum_{i=1}^{q} x_i y_l - \beta \sum_{i=1}^{q} x_l y_i + \beta^2 \sum_{i=1}^{q} x_i x_l \right] \right\} + \sum_{l=1}^{q}(y_l - \beta x_l)x_l \right] =$$

$$= \frac{1}{\sigma^2(1-\psi^2)} \left[ \sum_{i=1}^{q} x_i y_i - \beta \sum_{i=1}^{q} x_i^2 - \psi \sum_{i=1}^{q} x_i y_l - \psi \sum_{i=1}^{q} x_l y_i + 2\beta\psi \sum_{i=1}^{q} x_i x_l + \sum_{l=1}^{q} x_l y_l - \beta \sum_{l=1}^{q} x_l^2 \right] =$$

$$= \frac{1}{\sigma^2(1-\psi^2)} \left[ \sum_{i=1}^{q} x_i y_i + \sum_{l=1}^{q} x_l y_l - \psi \left( \sum_{i=1}^{q} x_i y_l + \sum_{l=1}^{q} x_l y_i \right) + \beta \left( 2\psi \sum_{i=1}^{q} x_i x_l - \sum_{i=1}^{q} x_i^2 - \sum_{l=1}^{q} x_l^2 \right) \right] \qquad (8)$$

which can be expressed in terms of the sufficient statistics (6) as

$$\frac{\partial}{\partial \beta} l(\beta, \sigma^2, \psi) = \frac{\alpha_3 - \beta\alpha_1 - \psi\alpha_4 + 2\beta\psi\alpha_5}{\sigma^2(1-\psi^2)} \qquad (9)$$

Assuming $|\psi| \neq 1$, and $\sigma^2 \neq 0$, and equating to zero Equation (9) we have:

$$\hat{\beta}_{BML} = \frac{\psi \left( \sum_{i=1}^{q} x_i y_l + \sum_{l=1}^{q} x_l y_i \right) - \left( \sum_{i=1}^{q} x_i y_i + \sum_{l=1}^{q} x_l y_l \right)}{\left( 2\psi \sum_{i=1}^{q} x_i x_l - \sum_{i=1}^{q} x_i^2 - \sum_{l=1}^{q} x_l^2 \right)}$$

or

$$\hat{\beta}_{BML} = \frac{\psi\alpha_4 - \alpha_3}{2\psi\alpha_5 - \alpha_1} \qquad (10)$$

The second first-order condition can be obtained from Equation (5) as:

$$\frac{\partial}{\partial \sigma^2} \lambda(\beta, \sigma^2, \psi) = \frac{\partial}{\partial \sigma^2} \left\{ const - q\ln(\sigma^2) - \frac{q}{2}\ln(1-\psi^2) - \frac{1}{2\sigma^2(1-\psi^2)} \sum_{i=1}^{q} \left[ \varepsilon_i^2 - 2\psi\varepsilon_i\varepsilon_I + \varepsilon_I^2 \right] \right\}$$

$$-\frac{q}{\sigma^2} + \frac{1}{2\sigma^4(1-\psi^2)} \sum_{i=1}^{q} \left[ \varepsilon_i^2 - 2\psi\varepsilon_i\varepsilon_I + \varepsilon_I^2 \right] \tag{11}$$

or, in terms of the sufficient statistics:

$$-\frac{q}{\sigma^2} + \frac{\alpha_2 - 2\beta\alpha_3 + \beta^2\alpha_1 - 2\beta\psi\alpha_4 - 2\beta^2\psi\alpha_5}{2\sigma^4(1-\psi^2)} \tag{12}$$

To obtain the estimator of $\sigma^2$, let us multiply both sides by $\sigma^4$ and equate (12) to zero. We have:

$$-q\sigma^2 + \frac{1}{2(1-\psi^2)} \sum_{i=1}^{q} \left[ \varepsilon_i^2 - 2\psi\varepsilon_i\varepsilon_I + \varepsilon_I^2 \right] = 0$$

hence

$$\hat{\sigma}^2_{BML} = \frac{1}{2q(1-\psi^2)} \sum_{i=1}^{q} \left[ \varepsilon_i^2 - 2\psi\varepsilon_i\varepsilon_I + \varepsilon_I^2 \right] \tag{13}$$

or, in terms of the sufficient statistics:

$$\hat{\sigma}^2_{BML} \frac{\alpha_2 + \beta^2\alpha_1 - 2\beta\alpha_3 - 2\psi\alpha_6 - 2\psi\beta^2\alpha_5 + 2\psi\beta\alpha_4}{2q(1-\psi^2)} \tag{14}$$

Finally, from Equation (5), the third first-order condition is given by:

$$\frac{\partial}{\partial \psi} l(\beta, \sigma^2, \psi) = \frac{\partial}{\partial \psi}\left\{ const - q\ln(\sigma^2) - \frac{q}{2}\ln(1-\psi^2) - \frac{\sum_{i=1}^{q}\varepsilon_i^2}{2\sigma^2(1-\psi^2)} + \frac{2\psi\sum_{i=1}^{q}\varepsilon_i\varepsilon_l}{2\sigma^2(1-\psi^2)} - \frac{\sum_{i=1}^{q}\varepsilon_l^2}{2\sigma^2(1-\psi^2)}\right\} =$$

$$= \frac{2q\psi}{2(1-\psi^2)} - \frac{2\psi\sum_{i=1}^{q}\varepsilon_i^2}{2\sigma^2(1-\psi^2)^2} - \frac{2\psi\sum_{i=1}^{q}\varepsilon_l^2}{2\sigma^2(1-\psi^2)^2} + \frac{2\sum_{i=1}^{q}\varepsilon_i\varepsilon_l}{2\sigma^2}\frac{\partial}{\partial \psi}\left\{\frac{\psi}{(1-\psi^2)}\right\} =$$

$$= \frac{q\psi}{(1-\psi^2)} - \frac{\psi\sum_{i=1}^{q}\varepsilon_i^2}{\sigma^2(1-\psi^2)^2} - \frac{\psi\sum_{i=1}^{q}\varepsilon_l^2}{\sigma^2(1-\psi^2)^2} + \frac{\sum_{i=1}^{q}\varepsilon_i\varepsilon_l}{\sigma^2}\frac{1+\psi^2}{(1-\psi^2)^2} \qquad (15)$$

or, expressed in terms of the sufficient statistics:

$$\frac{\partial}{\partial \psi} l(\beta, \sigma^2, \psi) =$$

$$\frac{q\psi}{(1-\psi^2)} - \frac{\psi(\alpha_2 - 2\beta\alpha_3 + \beta^2\alpha_1 - 2\psi\alpha_6 + 2\beta\psi\alpha_4 - 2\beta^2\psi\alpha_5)}{\sigma^2(1-\psi^2)^2} - \frac{-2\alpha_6 + 2\beta\alpha_4 - 2\beta^2\alpha_5}{2\sigma^2(1-\psi^2)} \qquad (16)$$

Equating Equation (15) to zero, multiplying both sides by $\sigma^2(1-\psi^2)^2$ (assuming again $|\psi|\neq 1$, and $\sigma^2 \neq 0$), we have a polynomial of order three. Two of the solutions have to be discarded ($\psi \pm 1$) while the third is:

$$\hat{\psi}_{BML} = \frac{\sum_{i=1}^{q}\varepsilon_i\varepsilon_l}{q\sigma^2} \qquad (17)$$

(For the purpose of illustration, Figure 2 displays the behavior of the loglikelihood as a function of ψ, and of the score function of ψ, for some artificial data).

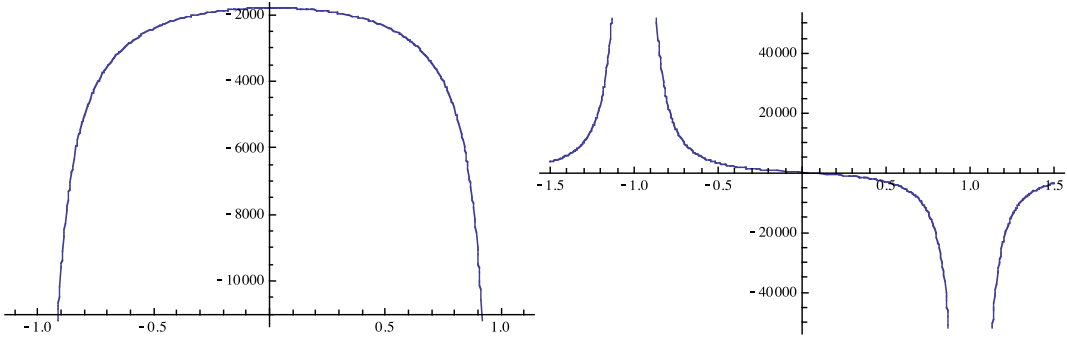

*Figure 1: Plot of $l(\psi)$ (left panel) and of $\frac{\partial}{\partial \psi} l(\psi)$ (right panel) in a set of artificial data when q=100, β=1, σ² = 1 and ψ =0.012.*

Finally, expressing (17) in terms of the sufficient statistics (6), we have:

$$\hat{\psi}_{BML} = \frac{\alpha_6 - \beta \alpha_4 + \beta^2 \alpha_5}{q\sigma^2} \quad (18)$$

Q.E.D.

Notice that, substituting in Equation (18) the *BML* solution obtained for $\sigma^2$ in equation (14), we have:

$$\hat{\psi}_{BML} = \frac{2(\alpha_6 - \beta \alpha_4 + \beta^2 \alpha_5)}{\alpha_2 - 2\beta \alpha_3 + \beta^2 \alpha_1} \quad (19)$$

So that the *BML* estimators of β and ψ are the solutions of the system of two equations:

$$\begin{cases} \hat{\beta}_{BML} = \dfrac{\psi \alpha_4 - \alpha_3}{2\psi \alpha_5 - \alpha_1} \\ \hat{\psi}_{BML} = \dfrac{2(\alpha_6 - \beta \alpha_4 + \beta^2 \alpha_5)}{\alpha_2 - 2\beta \alpha_3 + \beta^2 \alpha_1} \end{cases} \quad (20)$$

whereas the estimator of σ² is obtained by substituting the solutions derived in Equation (20) into Equation (14).

Notice also that, if $\psi = 0$ (case of pairwise bivariate spatial independence of the regression disturbances), in Equation (10) we have:

$$\hat{\beta}_{BML} = \frac{\psi \alpha_4 - \alpha_3}{2\psi \alpha_5 - \alpha_1} = \frac{\alpha_3}{\alpha_1} = \frac{\sum_{j=1}^{2q} x_j y_j}{\sum_{j=1}^{2q} x_j^2} = \hat{\beta}_{ML}$$

Similarly, if $\psi = 0$ in Equation (13) we have:

$$\hat{\sigma}^2_{BML} = \frac{1}{2q}\sum_{i=1}^{q}\left[\varepsilon_i^2 + \varepsilon_{i'}^2\right] = \frac{1}{2q}\sum_{j=1}^{2q}\varepsilon_j^2 = \hat{\sigma}^2_{ML}$$

which correspond to the familiar *ML* estimators of, respectively, $\beta$ and $\sigma^2$ in the case of independent errors.

Finally notice that the estimator $\hat{\psi}_{BML}$ derived in Equation (17) corresponds to the intuitive estimator of the spatial correlation among the disturbances.

In the remainder of this section we will establish the statistical properties of the BML estimators. Henceforth, to simplify the expressions, we will drop the subscript in the estimators' expressions thus setting $\hat{\beta}_{BML} = \hat{\beta}$, $\hat{\sigma}^2_{BML} = \hat{\sigma}^2$ and $\hat{\psi}_{BML} = \hat{\psi}$.

**Theorem 2:** *Under Assumption 1, Definitions 1 and 2 and Theorem 1, the inverse of Fisher sample information matrix associated to the BML estimators is given by:*

$$I(\theta)^{-1} = I(\beta, \sigma^2, \psi)^{-1} = \begin{pmatrix} \dfrac{\sigma^2(1-\psi^2)}{\alpha_1 - 2\psi\alpha_5} & 0 & 0 \\ 0 & \dfrac{\sigma^6(1-\psi^2)}{-2\psi\alpha_6 + 2\beta\psi\alpha_4 - 2\beta^2\psi\alpha_5 + q\sigma^2 + q\sigma^2\psi^2} & 0 \\ 0 & 0 & \dfrac{(1-\psi^2)^2}{q(1+\psi^2)} \end{pmatrix}$$

*Proof*: Let us first define the Hessian associated to the likelihood as $H_{ij}(\theta) = -\frac{\partial^2 l(\theta)}{\partial \theta \partial \theta^T}$; $i, j = 1, 2, 3$. By taking the derivatives of (9), (11) and (15) we obtain the elements of the hessian defined by:

$$H_{11} = \frac{\partial^2}{\partial \beta^2} l(\beta, \sigma^2, \psi) =$$

$$= \frac{1}{\sigma^2 (1 - \psi^2)} \left( \sum_{i=1}^{q} x_i^2 + \sum_{l=1}^{q} x_l^2 - 2\psi \sum_{l=1}^{q} x_i x_l \right) = \frac{\alpha_1 - 2\psi \alpha_5}{\sigma^2 (1 - \psi^2)} \quad (21)$$

$$H_{12} = \frac{\partial^2}{\partial \beta \partial \sigma^2} l(\beta, \sigma^2, \psi) =$$

$$-\frac{1}{\sigma^4 (1 - \psi^2)} \left[ \sum_{i=1}^{q} x_i y_i + \sum_{l=1}^{q} x_l y_l - \psi \left( \sum_{i=1}^{q} x_i y_l + \sum_{l=1}^{q} x_l y_i \right) + \beta \left( 2\psi \sum_{l=1}^{q} x_i x_l - \sum_{i=1}^{q} x_i^2 - \sum_{l=1}^{q} x_l^2 \right) \right] =$$

$$= \frac{\alpha_3 - \beta \alpha_1 - \psi \alpha_4 - 2\beta \psi \alpha_5}{\sigma^4 (1 - \psi^2)} \quad (22)$$

$$H_{13} = \frac{\partial^2}{\partial \beta \partial \psi} l(\beta, \sigma^2, \psi) =$$

$$\frac{\partial}{\partial \psi} \frac{1}{\sigma^2 (1 - \psi^2)} \left[ \sum_{i=1}^{q} x_i y_i + \sum_{l=1}^{q} x_l y_l - \psi \left( \sum_{i=1}^{q} x_i y_l + \sum_{l=1}^{q} x_l y_i \right) + \beta \left( 2\psi \sum_{l=1}^{q} x_i x_l - \sum_{i=1}^{q} x_i^2 - \sum_{l=1}^{q} x_l^2 \right) \right] =$$

$$= -\frac{2\psi}{\sigma^2 (1 - \psi^2)^2} \left[ \sum_{i=1}^{q} x_i y_i + \sum_{l=1}^{q} x_l y_l - \psi \left( \sum_{i=1}^{q} x_i y_l + \sum_{l=1}^{q} x_l y_i \right) + \beta \left( 2\psi \sum_{l=1}^{q} x_i x_l - \sum_{i=1}^{q} x_i^2 - \sum_{l=1}^{q} x_l^2 \right) \right] +$$

$$+ \frac{1}{\sigma^2 (1 - \psi^2)} \left[ -\sum_{i=1}^{q} x_i y_l - \sum_{l=1}^{q} x_l y_i + 2\beta \sum_{l=1}^{q} x_i x_l \right] =$$

$$= -\frac{\psi(-2\alpha_3 + 2\beta \alpha_1 + 2\psi \alpha_4 - 4\beta \psi \alpha_5)}{\sigma^2 (1 - \psi^2)^2} - \frac{\alpha_4 - 2\beta \alpha_5}{\sigma^2 (1 - \psi^2)} \quad (23)$$

$$H_{22} = \frac{\partial^2}{\partial(\sigma^2)^2} l(\beta,\sigma^2,\psi) = \frac{q}{\sigma^4} - \frac{2\sigma^2}{2\sigma^8(1-\psi^2)}\sum_{i=1}^{q}\left[\varepsilon_i^2 - 2\psi\varepsilon_i\varepsilon_I + \varepsilon_I^2\right] =$$

$$\frac{q}{\sigma^4} - \frac{1}{\sigma^6(1-\psi^2)}\sum_{i=1}^{q}\left[\varepsilon_i^2 - 2\psi\varepsilon_i\varepsilon_I + \varepsilon_I^2\right]$$

$$= \frac{q}{\sigma^4} - \frac{\alpha_2 + \beta^2\alpha_1 - 2\beta\alpha_3 - 2\psi\alpha_6 + 2\beta\psi\alpha_4 - 2\beta^2\psi\alpha_5}{\sigma^6(1-\psi^2)} \qquad (24)$$

$$H_{23} = \frac{\partial^2}{\partial\psi\partial\sigma^2} l(\beta,\sigma^2,\psi) = \frac{\partial}{\partial\psi}\left\{-\frac{q}{\sigma^2} + \frac{1}{2\sigma^4(1-\psi^2)}\sum_{i=1}^{q}\left[\varepsilon_i^2 - 2\psi\varepsilon_i\varepsilon_I + \varepsilon_I^2\right]\right\} =$$

$$= \frac{\partial}{\partial\psi}\left\{-\frac{q}{\sigma^2} + \frac{\sum_{i=1}^{q}\varepsilon_i^2}{2\sigma^4(1-\psi^2)} - \frac{2\psi\sum_{i=1}^{q}\varepsilon_i\varepsilon_I}{2\sigma^4(1-\psi^2)} + \frac{\sum_{i=1}^{q}\varepsilon_I^2}{2\sigma^4(1-\psi^2)}\right\} =$$

$$= \frac{2\psi\sum_{i=1}^{q}\varepsilon_i^2}{2\sigma^4(1-\psi^2)^2} + \frac{2\psi\sum_{i=I}^{q}\varepsilon_I^2}{2\sigma^4(1-\psi^2)^2} - \frac{2\sum_{i=1}^{q}\varepsilon_i\varepsilon_I}{2\sigma^4}\frac{\partial}{\partial\psi}\frac{\psi}{(1-\psi^2)} =$$

$$= \frac{\psi\sum_{i=1}^{q}\varepsilon_i^2}{\sigma^4(1-\psi^2)^2} + \frac{\psi\sum_{i=I}^{q}\varepsilon_I^2}{\sigma^4(1-\psi^2)^2} - \frac{1+\psi^2}{(1-\psi^2)^2}\frac{\sum_{i=1}^{q}\varepsilon_i\varepsilon_I}{\sigma^4} =$$

$$= \frac{\psi(\alpha_2 - 2\beta\alpha_3 + \beta^2\alpha_1 - 2\psi\alpha_6 - 2\beta\psi\alpha_4 - 2\beta^2\alpha_5)}{\sigma^4(1-\psi^2)^2} - \frac{2\alpha_6 - 2\beta\alpha_4 + 2\beta^2\alpha_5}{2\sigma^4(1-\psi^2)^2} \qquad (25)$$

and, finally,

$$H_{33} = \frac{\partial^2}{\partial\psi^2} l(\beta,\sigma^2,\psi) =$$

$$\frac{\partial}{\partial \psi}\left[-\frac{q\psi}{(1-\psi^2)} - \frac{\psi\sum_{i=1}^{q}\varepsilon_i^2}{\sigma^2(1-\psi^2)^2} - \frac{\psi\sum_{l=1}^{q}\varepsilon_l^2}{\sigma^2(1-\psi^2)^2} + \frac{\sum_{i=1}^{q}\varepsilon_i\varepsilon_l}{\sigma^2}\frac{1+\psi^2}{(1-\psi^2)^2}\right]=$$

$$=-q\frac{(1+\psi^2)}{(1-\psi^2)^2} - \left[\frac{\sum_{i=1}^{q}\varepsilon_i^2}{\sigma^2} + \frac{\sum_{l=1}^{q}\varepsilon_l^2}{\sigma^2}\right]\frac{(1+3\psi^2)}{(1-\psi^2)^3} + \frac{\sum_{i=1}^{q}\varepsilon_i\varepsilon_l}{\sigma^2}\left[\frac{2\psi(\psi^2+3)}{(1-\psi^2)^3}\right] =$$

$$\frac{2q\psi^2}{(1-\psi^2)^2} + \frac{q}{(1-\psi^2)} - \frac{\alpha_2 - 2\beta\alpha_1 + \beta^2\alpha_1 - 2\psi\alpha_6 + 2\beta\psi\alpha_4 - 2\beta^2\psi\alpha_5}{\sigma^2(1-\psi^2)^3} -$$

$$-\frac{\alpha_2 - 2\beta\alpha_1 + \beta^2\alpha_1 - 2\psi\alpha_6 + 2\beta\psi\alpha_4 - 2\beta^2\psi\alpha_5}{\sigma^2(1-\psi^2)^3}$$

$$-\frac{2\psi\left(-2\alpha_6\alpha_2 + 2\beta\alpha_4 - 2\beta^2\alpha_5\right)}{\sigma^2(1-\psi^2)^2} - \frac{\alpha_2 - 2\beta\alpha_3 + \beta^2\alpha_1 - 2\psi\alpha_6 + 2\beta\psi\alpha_4 - 2\beta^2\psi\alpha_5}{\sigma^2(1-\psi^2)^2} \tag{26}$$

Let us now derive the elements of Fisher's sample Information matrix $I(\theta) = E[H(\theta)]$. First of all consider that, from *Definition 1*, we have the following results:

$$E(\alpha_1) = \alpha_1 \qquad E(\alpha_2) = \beta^2\alpha_1 \qquad E(\alpha_3) = \beta\alpha_1$$
$$E(\alpha_4) = 2\beta\alpha_5 \qquad E(\alpha_5) = \alpha_5 \qquad E(\alpha_6) = \beta^2\alpha_5$$

As a consequence we have that the element of Fisher's Information matrix are given by:

$$I_{11} = -E(H_{11}) = \frac{\alpha_1 - 2\psi\alpha_5}{\sigma^2(1-\psi^2)}$$

$$I_{12} = -E(H_{12}) = 0$$

$$I_{13} = -E(H_{13}) = 0$$

$$I_{22} = -E(H_{22}) = \frac{-2\psi\alpha_6 + 2\beta\psi\alpha_4 - 2\beta^2\psi\alpha_5}{\sigma^6(1-\psi^2)} + \frac{q}{\sigma^4}$$

$$I_{23} = -E(H_{23}) = 0$$

$$I_{33} = -E(H_{33}) = \frac{q(1+\psi^2)}{(1-\psi^2)^2}$$

Finally the elements of the inverse of Fisher expected information matrix are equal to:

$$I_{11}^{-1} = \frac{\sigma^2(1-\psi^2)}{\alpha_1 - 2\psi\alpha_5} \tag{27}$$

$$I_{22}^{-1} = \frac{\sigma^6(1-\psi^2)}{-2\psi\alpha_6 + 2\beta\psi\alpha_4 - 2\beta^2\psi\alpha_5 + q\sigma^2 + q\sigma^2\psi^2} \tag{28}$$

$$I_{33}^{-1} = \frac{(1-\psi^2)^2}{q(1+\psi^2)} \tag{29}$$

and $I_{12}^{-1} = I_{13}^{-1} = I_{23}^{-1} = 0$.

Q.E.D.

**Corollary 1:** The BML estimators of the the model's parameters β, σ² and ψ are mutually incorrelated.

The corollary immediately follows observing that the information matrix $I(\beta, \sigma^2, \psi)^{-1}$ is diagonal.

**Theorem 3:** Under Assumption 1 and Theorem 1, $\hat{\beta}_{BMML}$, $\hat{\sigma}^2_{BMML}$ and $\hat{\psi}_{PML}$ are unbiased estimators of the model's parameters β, σ² and ψ.

*Proof*: To start with, from Equation (10) we have:

$$E(\hat{\beta}) = \frac{\psi\left(\sum_{i=1}^{q} x_i y_l + \sum_{l=1}^{q} x_l y_i\right) - \left(\sum_{i=1}^{q} x_i y_i + \sum_{l=1}^{q} x_l y_l\right)}{2\psi\sum_{l=1}^{q} x_i x_l - \sum_{i=1}^{q} x_i^2 - \sum_{l=1}^{q} x_l^2}$$

and, since x is non stochastic, for any given value of $\hat{\psi}$ we have:

$$E(\hat{\beta}) = \frac{\sum_{i=1}^{q} x_i E(y_i) + \sum_{l=1}^{q} x_l E(y_l) - \hat{\psi}\left(\sum_{l=1}^{q} x_l E(y_l) + \sum_{i=1}^{q} x_i E(y_i)\right)}{\sum_{i=1}^{q} x_i^2 + \sum_{l=1}^{q} x_l^2 - 2\hat{\psi}\sum_{l=1}^{q} x_i x_l}$$

From Assumption 1, conditionally on $\hat{\psi}$, the expected value of y is $E(y_i|\hat{\psi}) = E(\beta x_i + \varepsilon_i|\hat{\psi})$, and, since x is non stochastic and $E(\varepsilon_i|\hat{\psi}) = 0$, then $E(y_i|\hat{\psi}) = \beta x_i$. Hence:

$$E(\hat{\beta}) = \frac{\sum_{i=1}^{q} x_i \beta x_i + \sum_{l=1}^{q} x_l \beta x_l - \hat{\psi}\left(\sum_{i=1}^{q} x_i \beta x_l + \sum_{l=1}^{q} x_l \beta x_i\right)}{\sum_{i=1}^{q} x_i^2 + \sum_{l=1}^{q} x_l^2 - 2\hat{\psi}\sum_{l=1}^{q} x_i x_l} =$$

$$= \beta \frac{\sum_{i=1}^{q} x_i^2 + \sum_{l=1}^{q} x_l^2 - \hat{\psi}\left(\sum_{i=1}^{q} x_i x_l + \sum_{l=1}^{q} x_l x_i\right)}{\sum_{i=1}^{q} x_i^2 + \sum_{l=1}^{q} x_l^2 - 2\hat{\psi}\sum_{l=1}^{q} x_i x_l} = \beta$$

which proves unbiasedness.

Q.E.D.

**Theorem 4:** *Under Assumption 1, Definitions 1 and 2 and Theorem 1, $\hat{\beta}$ is normally distributed.*

*Proof*: From Assumption 1 it follows that $\varepsilon_i \approx N(0, \sigma^2)$. Hence, since x is non stochastic from Equation (1), we also have $y_i \approx N(0, \sigma^2)$. As a consequence $\alpha_3 = \sum_{i=1}^{q} x_i y_i + \sum_{l=1}^{q} x_l y_l = \sum_{j=1}^{2q} x_j y_j$ and $\alpha_4 = \sum_{i=1}^{q} x_i y_l + \sum_{l=1}^{q} x_l y_i$ are also normally distributed, while $\alpha_1 = \sum_{i=1}^{q} x_i^2 + \sum_{l=1}^{q} x_l^2$ and $\alpha_5 = \sum_{l=1}^{q} x_i x_l$ are non stochastic. So recalling Equation (10) we have that, conditionally on the value of $\psi$, $\hat{\beta} \approx N$, with expected value $E(\hat{\beta}) = \beta$ as proved in Theorem 3 and with $Var(\hat{\beta}) = \frac{\sigma^2(1-\psi^2)}{\alpha_1 - 2\psi\alpha_5}$ as proved in Theorem 2 (Equation (27)).

Q.E.D.

**Theorem 5:** *Under Assumption 1, Definitions 1 and 2, Theorem 2 and Theorem 3, $\hat{\beta}$ is a weakly consistent estimator of the model's parameter β.*

*Proof:* In Theorem 3 we proved that $\hat{\beta}$ is an unbiased estimator. Thus a sufficient condition for weak consistency is $\lim_{q \to \infty} Var(\hat{\beta}_{BMML}) = 0$, $q$ being (half of the) sample size. From Theorem 2 we have:

$$Var(\hat{\beta}) = \frac{\hat{\sigma}^2 (1 - \hat{\psi}^2)}{\alpha_1 - 2\hat{\psi}\alpha_5} \tag{30}$$

If in equation (30) we substitute the explicit solution for $\hat{\sigma}^2$ obtained in Equation (14), we have:

$$Var(\hat{\beta}) = \frac{\alpha_2 + \hat{\beta}^2 \alpha_1 - 2\hat{\beta}\alpha_3 - 2\hat{\psi}\alpha_6 - 2\hat{\psi}\hat{\beta}^2 \alpha_5 + 2\hat{\psi}\hat{\beta}\alpha_4}{2q(\alpha_1 - 2\hat{\psi}\alpha_5)}$$

and this expression tends to zero for $q \to \infty$ thus proving the second part of the theorem.

Q.E.D.

In Equation (1) the assumption that $x_i$ is a scalar can be easily generalized to allow for a vector of predictors. Consider now the following regression model

$$y = X\beta + \varepsilon \tag{31}$$

with y a *2q-by-1* vector, X a *2q-by-k* matrix of observations of *k* predictors, β a *k-by-1* vector of parameters and ε a *2q-by-1* vector of disturbances. Let us now partition the matrices appearing in Equation (31) according to the following definitions.

**Definition 3:** *Define the following matrices:*

$$y^{(1)} = \begin{bmatrix} y_1^{(1)} \\ \ldots \\ y_i^{(1)} \\ \ldots \\ y_q^{(1)} \end{bmatrix}; \; y^{(2)} = \begin{bmatrix} y_1^{(2)} \\ \ldots \\ y_l^{(2)} \\ \ldots \\ y_q^{(2)} \end{bmatrix}; \; y = \begin{bmatrix} y^{(1)} \\ y^{(2)} \end{bmatrix} \quad (32)$$

$$X^{(1)} = \begin{bmatrix} {}_1x_1^{(1)} & {}_2x_1^{(1)} & & {}_kx_1^{(1)} \\ {}_1x_i^{(1)} & {}_2x_i^{(1)} & & {}_kx_i^{(1)} \\ {}_1x_q^{(1)} & {}_2x_q^{(1)} & & {}_kx_q^{(1)} \end{bmatrix}; \; X^{(2)} = \begin{bmatrix} {}_1x_1^{(2)} & {}_2x_1^{(2)} & & {}_kx_1^{(2)} \\ {}_1x_l^{(2)} & {}_2x_l^{(2)} & & {}_kx_l^{(2)} \\ {}_1x_q^{(2)} & {}_2x_q^{(2)} & & {}_kx_q^{(2)} \end{bmatrix}; \; X = \begin{bmatrix} X^{(1)} \\ X^{(2)} \end{bmatrix} \quad (33)$$

and

$$\varepsilon^{(1)} = \begin{bmatrix} \varepsilon_1^{(1)} \\ \ldots \\ \varepsilon_i^{(1)} \\ \ldots \\ \varepsilon_q^{(1)} \end{bmatrix}; \; \varepsilon^{(2)} = \begin{bmatrix} \varepsilon_1^{(2)} \\ \ldots \\ \varepsilon_l^{(2)} \\ \ldots \\ \varepsilon_q^{(2)} \end{bmatrix}; \; \varepsilon = \begin{bmatrix} \varepsilon^{(1)} \\ \varepsilon^{(2)} \end{bmatrix} \quad (34)$$

In the previous expressions the superscript (1) refers to the first element and the superscript (2) to the second element of the same selected units in such a way that $\forall j = 1, \ldots, q$, $y_j^{(1)}$ and $y_j^{(2)}$ are two neighboring observations of the variable y and similarly for the variables $x$ and $\varepsilon$.

***Theorem 6:*** *Given Assumption 1 and Definitions 3, by extension of the result contained in Theorem 1, the bivariate marginal maximum likelihood estimator of the vector of parameters $\theta = (\beta, \sigma^2, \psi)$, , are given by the solutions of the following system of equations:*

$$\hat{\beta} = \left[ X^{(1)T}X^{(1)} + X^{(2)T}X^{(2)} + 2\hat{\psi}X^{(1)T}X^{(2)} \right]^{-1} \left[ X^{(1)T}y^{(1)} + X^{(2)T}y^{(2)} + \hat{\psi}\left( X^{(1)T}y^{(2)} + X^{(2)T}y^{(1)} \right) \right] \quad (35)$$

$$\hat{\sigma}^2 = \frac{\varepsilon^{(1)T}\varepsilon^{(1)} - 2\hat{\psi}\varepsilon^{(1)T}\varepsilon^{(2)} + \varepsilon^{(2)T}\varepsilon^{(2)}}{2q(1-\hat{\psi}^2)} \quad (36)$$

$$\hat{\psi} = \frac{\varepsilon^{(1)T}\varepsilon^{(2)}}{q\hat{\sigma}^2} \tag{37}$$

where $\hat{\beta}$ is a *k-by-1* vector and $\hat{\sigma}^2$ and $\hat{\psi}$ are scalar.

### 3. Interpretation of the parameters

Apart from the analytical advantages of the spatial econometric model illustrated in the previous section, the Bivariate Maximum Likelihood Estimators proposed in this paper also provides some interpretative advantages. First of all, in our setting, the parameter β can be interpreted in the traditional way as the impact on variable y of an infinitesimal change in variable x $\left(\beta = \frac{\partial y_i}{\partial x_i}\right)$. In this respect our specification avoids the more complicated interpretations associated with the traditional SARAR frameworks where "a change in the explanatory variable for a single region (observation) can potentially affect the dependent variable in all other observations" (LeSage and Pace, 2010, p. 35). See also the discussion on the spatial multipliers contained in Kelejian et al. (2006). Furthermore, consider the formal expression of the BML estimator of β derived in Theorem 1 and reported here for convenience $\hat{\beta}_{BML} = \frac{\alpha_3 - \psi\alpha_4}{\alpha_1 - 2\psi\alpha_5}$. The numerator of this expression represents by the covariance between x and y (the term $\alpha_3$) augmented with the extra term $-\psi\alpha_4$ which represents the spatial spill-over of the variable x in one location onto the variable y in a neighboring location belonging to the same coding unit (the term $\alpha_4$), weighted with the disturbance spatial correlation parameter ψ. Similarly the denominator represents the variance of the independent variable (the term $\alpha_1$) augmented with the term $-2\psi\alpha_5$ representing the spatial autocovariance of variance x (the term $\alpha_5$), weighted with the spatial correlation of the error term.

The interpretation of the parameter is rather straightforward. Consider, e. g., the case of positive error spatial correlation ($\psi > 0$). If the spatial spill-over between *x* and *y* and the variance of *x* are of different sign, the multiplicative effect of *x* on *y* is affected. In particular, the effect will be emphasized if the spatial spillover ($\alpha_4$) is negative and the spatial autocovariance of *x* ($\alpha_5$) is positive and vice-versa. When $\alpha_4$ and $\alpha_5$ are of the same sign, in contrast, the effect on $\hat{\beta}_{BML}$ is more complex to analyze in that it depends on the absolute value of the sufficient statistics and on the value of $\psi$. All previous relationships are are reversed in the case of a negative error spatial correlation. Consider, as an example, the case of $\psi > 0$. In this case intuition suggests that the effect on y of a variation in the independent variable should be more pronounced in the presence of a positive spill-over between the two variables in that one location benefits not only from increases of x in the same location, but also for the increase of x in the neighboring locations. However the formal expression of $\hat{\beta}_{BML}$ shows that this is true only in the presence of a strong positive spatial correlation of the independent variable. Similarly the formal expression of the *BML* estimator of β also shows that we can have a higher impact of the independent variable on the dependent variable even if there is no spatial spill-over between the two. In fact, when $\alpha_4 = 0$, the effect on y of a variation in the independent variable will be more pronounced if $\alpha_5 > 0$, that is in the presence of a positive spatial correlation in the independent variable.

## 4. Final comments and future extensions

This paper introduces a new modeling framework for spatial error econometric models which allows the maximum likelihood solution to be expressed in closed form thus dramatically improving the numerical performances with respect to the more traditional spatial autoregressive model specifications. In this paper, to go straight to the point and to reduce the complexity, we have restricted ourselves to cross-sectional spatial linear models, but the framework described here can be extended to different specifications. A first immediate extension of the bivariate marginal approach refers to models of the class SARAR (Kelejian and Prucha, 1998) in the presence of both a spatial error and a spatial lag component. A second extension refers to spatial discrete choice models and to spatial panels. When dealing with nonlinear discrete choice models, the computational difficulties are even more pronounced than with linear models and the alternatives to the traditional ML estimators (like e.g. GMM solutions) do not entirely eliminate them as documented e. g. in Smirnov (2010). Similarly when we consider spatial panel data models both the quasi-likelihood and partial likelihood discussed in Lee and Yu (2010) and the GMM approach suggested by Kapoor *et al.* (2007) present serious computational issues especially when the spatial dimension of the dataset is very large. In this sense an extension of the procedure suggested in this paper would be of tremendous benefit in these two fields. Due to the results obtained in Theorem 2, standard likelihood-based hypothesis testing procedures can be applied to a bivariate marginal specification of a spatial econometric model. However, our framework also allows a further approach to standard errors evaluation and to hypothesis testing based on the idea of resampling. Resampling methods for sets of dependent random variables have a long tradition in statistics dating back to the earlier contributions of Solow (1985) and Arbia (1990) among the others. The bivariate coding technique presented here is based on the identification of a subsample of pairs of units to be used in a Bivariate Marginal Maximum Likelihood approach. However, such a coding scheme is non unique. Even in the simple example reported in Figure 1 we have that the procedure could lead to 4 different codings (leading to four different estimations of the model's parameters) corresponding to the first cross-coded unit being located in the first or the second cell in the first row or in the first and in the second cell in the second row. The number of possible configurations could be even larger in non-lattice irregular spatial data sets. This problem was already noted by Besag (1974) and the solution suggested was to average the estimates obtained under the different coding systems. In this paper, in the spirit of Besag's original proposal, we considered an exhaustive coverage of all the units, however, if we select a smaller number of coded units without exhaustively covering the whole space, we can consider many possible bivariate coding schemes, and correspondently we could obtain many different parameters' estimations allowing the derivation of a resampling distribution. Under this respect the approach presented here suggests a formal way of bootstrapping spatial data in a regression context preserving both the condition of independence between the subsamples and the spatial dependence features of the data.